\documentclass[english,aps,prl,showpacs,preprintnumbers,amsmath,amssymb,twocolumn]{revtex4-1}
\usepackage[T1]{fontenc}
\usepackage[latin9]{inputenc}
\usepackage{amsmath}
\usepackage{amssymb}
\usepackage{graphicx}

\makeatletter

\@ifundefined{textcolor}{}
{%
 \definecolor{BLACK}{gray}{0}
 \definecolor{WHITE}{gray}{1}
 \definecolor{RED}{rgb}{1,0,0}
 \definecolor{GREEN}{rgb}{0,1,0}
 \definecolor{BLUE}{rgb}{0,0,1}
 \definecolor{CYAN}{cmyk}{1,0,0,0}
 \definecolor{MAGENTA}{cmyk}{0,1,0,0}
 \definecolor{YELLOW}{cmyk}{0,0,1,0}
 }

\def\tev{\; \hbox{TeV}}
\def\gev{\; \hbox{GeV}}

\usepackage{amsfonts}
\usepackage{epsfig}\usepackage{dcolumn}
\usepackage{bm}
\date{\today}

\makeatother

\usepackage{babel}
\begin{document}

\title{Jets Plus Same-Sign Dileptons Signatures from Fermionic Leptoquarks
at the LHC
\bigskip{}
}

\author{Alexandre Alves$^{1}$, E. Ramirez Barreto$^{2}$, A. G. Dias$^{2}$\medskip{}
}

\affiliation{$^{1}$Universidade Federal de S\~ao Paulo, UNIFESP,  Dep. de Ci\^encias Exatas e da Terra,  Diadema - SP 09972-270 Brasil, \\
$^{2}$Universidade Federal do ABC, UFABC, Centro de Ci\^encias Naturais e Humanas,  Santo Andr\'e - SP, 09210-170, Brasil.}


\begin{abstract}

Within the 3-3-1 model framework, we consider the production and decay of fermionic leptoquarks into a striking
 experimental signature at the LHC: a narrow resonance at the $b$ jet plus same-sign dileptons channel. The data already accumulated by the LHC collaborations may hide a large number of events associated to the production and decay of such exotic  quarks, allowing one to investigate large portions of the parameters space of the model. The observation or not of such events would shed light on the construction of models presenting heavy exotic quarks.

\end{abstract}
\maketitle

The  hypothesis of the existence of new particles carrying two units of lepton number
allows us to consider new physical phenomena which can be quite notable, if their masses are within the actual testable TeV energy scale. The decay of such exotic particles would lead to signatures that can not arise from any novel heavy states whose quantum numbers are the same as the ones in the standard model (SM). As a result, the signals would be almost free from SM backgrounds.

Let us consider the pair production process of \emph{fermionic leptoquarks}, $J$, that carry both lepton and baryon number $(L,B)=(2,1/3)$. If all interactions preserve these quantum numbers, the new quark may decay into a vector (or scalar) $(L,B)=(2,0)$ \emph{bilepton} plus a fermionic $(L,B)=(0,1/3)$ particle. For a SM quark in the final state, the vector (or scalar) bilepton is cons\-tra\-ined to decay into: (1) a pair of same-sign charged leptons, or (2) a charged lepton and a neutrino through a fermion-flow-violating interaction. General information on scalar and vector bileptons/leptoquarks can be found in Refs.~\cite{CUY}, \cite{PDG}.


Under the above assumptions, a new colored heavy state could be realized either with a electric charge $-4/3$ or $5/3$ (in units of
 the electron charge, $e$). The corresponding bileptons would have electric charge $1$ and $2$. We call such bileptons of $V$ and $U$,
 respectively. Now, if we discard model-dependent Yukawa interactions and allow just gauge interactions, the bileptons are
spin-1 particles.

This way, the fermionic leptoquark decays by means of charged currents to a down- or up-type quark plus a $U$ or $V$. Limiting
the effects of the new sector to the third-generation SM quarks and leptoquarks $J_3$, there are four  decay possibilities:
\begin{eqnarray}
\label{eq:channels}
J_3(+5/3) & \rightarrow & U^{++} + b(-1/3) \rightarrow \ell^+\ell^++b \\
J_3(-4/3) & \rightarrow & U^{--} + t(+2/3)\rightarrow \ell^-\ell^-+t \\
J_3(+5/3) & \rightarrow & V^{+\; } + t(+2/3) \rightarrow \ell^+\overline{\nu}_\ell+t \\
J_3(-4/3) & \rightarrow & V^{-\; } + b(-1/3) \rightarrow \ell^-\nu_\ell+b
\end{eqnarray}
In the third and fourth channels, the experimental signatures would be similar to a heavy top or bottom quark partner decay
 as proposed in new strong dynamics models or fourth-generation quark models~\cite{alternatives}, for example. However,
in such models, $V$ is not a bilepton, so  the searching for the doubly charged bileptons, from the first and second
 channels, would be a necessary step to decide which model would be favored by the data.

A specific theoretical realization of leptoquarks and bileptons we deal with in this work arise as a part of the particle
 spectrum of SM extensions based on the
$SU(3)_C\otimes SU(3)_L\otimes U(1)_X$ gauge group, the 3-3-1 models. Within this class of models, the minimal 3-3-1 model
 ~\cite{331ppf} is a natural basis for constructions presenting a leptoquark $J_3$ and bileptons $U$ and $V$. These models
 are also motivated by offering an explanation for the families number problem, and  for constraining the electroweak
 mixing angle such that $\sin\theta_W < 0.25$. The phenomenology of the 3-3-1 models aiming the colliders has been developed
through the years~\cite{Pheno}. Another interesting feature of the model is an enhanced Higgs boson decay into photons that could explain the recent results of the experiments at the LHC concerning a  Higgs boson with mass around $\sim 125$ GeV, as shown in~\cite{h2f331min,h125331}.

We show in this letter that within the minimal 3-3-1 model framework, a large number of events associated to the production and decay of fermionic leptoquarks might have been produced at the 7 TeV LHC with an almost negligible level of background noise. We also investigate the potential of the 14 TeV LHC to extend the discovery reach for fermionic leptoquarks of the 3-3-1 model. The current amount of collected data would render the discovery of fermionic leptoquarks and doubly charged bileptons an easy task already, even for masses of order of TeV. If no events can be found, a prediction consistent with the SM expectation, large portions of the parameters space of the proposed 3-3-1 model can be ruled out at a high confidence level.

Moreover, ruling out the 3-3-1 model is a necessary step to discriminate between alternative exotic quark models, for example, new strong dynamics, grand unified, superstring, or fourth-generation models~\cite{alternatives}. As a final remark, we point out that existing bounds
 on scalar and vector leptoquarks~\cite{PDG} do not apply to a fermionic leptoquark.

It has to be said that there are few other models containing fermionic leptoquarks \cite{other-lept}.  But none of these models the leptoquarks  are coupled to doubly charged vector bosons. In fact,  such  coupling is what leads to the distinct process we are proposing here.

In what follows, we present briefly the essential aspects of
the 3-3-1 models. Then, we analyze  the pair production and decay signal of leptoquarks, and  finish with a
discussion of our results.

\section{Brief Review of the Model}

 The matter content of 3-3-1 models is defined in such a way that gauge anomalies
 cancellation of the group $SU(3)_C\otimes SU(2)_L\otimes U(1)_Y$ involves all three fermionic families. Different versions
of the model can be constructed with this restriction. The  matter content of the minimal 3-3-1 model has three lepton triplets,
 each one formed by the left-handed field of a neutrino, a charged lepton, and its antiparticle: $\Psi_{a}\equiv(\nu_{a}^\prime \,\, l_{a}^\prime\,\, l_{a}^{^\prime c})_L\sim\left(\mathbf{3}-1/3\right)$, where $a=1,2,3$
is the family index; and the numbers in parentheses
refer to the trans\-for\-ma\-tion character under $SU(3)_{L}$ and $U(1)_{X}$,
respectively (we will omit  the color group $SU(3)_C$ transformation character of the multiplets). The primes means the fields
 are not mass
eigenstates. For the quarks, they form  two antitriplets
$Q_{n}\equiv\left (d_n^\prime\,,-u_n^\prime \,, j_n^\prime \right )_L \sim({{{\bf 3^*}}},-\frac{1}{3})$, and a triplet
 $Q_{3}\equiv \left (u_3^\prime \,,  d_3^\prime \,, J_3\right )_L \sim ({{\bf 3}},\frac{2}{3})$ of left-handed fields; here,
the indices $n=1,2$ and $3$ refer to the usual quark fa\-mi\-lies, with the right-handed quarks fields forming the singlets
 $u_{i_R}^\prime\sim(\mbox{{\bf 1}},\frac{2}{3})$, $d_{i_R}^\prime\sim(\mbox{{\bf 1}},-\frac{1}{3})$,
 $j_{n_R}^\prime\sim(\mbox{{\bf 1}},
-\frac{4}{3})$, $J_{3R}\sim(\mbox{{\bf 1}},\frac{5}{3})$, where $i=1,2,3$,   $j_{n_R}^\prime$, and $J_3$ are new quarks with
 electric charge $-4/3$ and $5/3$, respectively. $J_3$ is the fermionic leptoquark we focus on this work.

Spontaneous symmetry breaking and mass generation can be implemented with three scalar triplets plus a sextet~\cite{331ppf},
or with just two scalar triplets~\cite{331minesc}. We refer the readers to those references for further details concerning this point.

Besides the SM gauge bosons, the model has five additional massive states: a neutral $Z^{\prime}$, and the  electrically charged
bileptons, $V$ and  $U$.

For our purposes the relevant new charged current interactions for the mass eigenstates  are the following: one which involves
the leptoquark $J_3$, the bottom quark $b$, and the bilepton $U^{\pm \pm}$
\begin{equation}
\mathcal{L}_{bJU}=-\frac{g}{2\sqrt{2}}\overline{b}\gamma^{\mu}(1-\gamma^5)J_3\, U_{\mu}^{--}+H.c.,
\label{jbu}
\end{equation}
where $g$ is the $SU(2)_L$ gauge coupling constant; and another one which involves the same bileptons and charged lepton fields
in a purely pseudovectorial fermion-flow-vio\-la\-ting interaction
\begin{equation}
\mathcal{L}_{\ell\ell U}=-\frac{g}{2\sqrt{2}}\overline{\ell^{c}_a}\gamma^{\mu}\gamma^{5}\ell_a\, U_{\mu}^{++}+H.c.
\label{lclu}
\end{equation}
Equations (\ref{jbu}) and (\ref{lclu}) are the interactions leading to the ge\-ne\-ral decay in Eq. (\ref{eq:channels}) and shown in Fig. \ref{fig:1}.

An explanation is due now. The weak eigenstates $b^\prime,s^\prime,d^\prime$ are linear combinations of the mass eigenstates $b,s,d$, so the branching ratio of the decay $J_3\rightarrow bU$ is, in fact, proportional to $|V^\prime_{bb}|^2$, the square of an element of the rotation matrix of the down sector including the new quark. However, we are not demanding the flavor of jets to be identified. Our partonic simulation is not affected if we either fix $V^\prime_{bb}=1$ and set the other elements to zero, or  assume nonzero values for the other $V^\prime$ elements, as long as we do not require any jet tagging.

In the leptoquarks production processes, besides the QCD, electromagnetic, and the neutral current interaction involving the equivalent of the SM $Z$ boson, there are additional contributions coming from new neutral current interactions  due to the $Z^\prime$.
 In order to simplify  our analysis, we assume that the vacuum expectation values of the scalar fields, taken to break the symmetries, obey a certain relation such that all neutral current couplings of known particles are identical to the SM as presented in Ref. \cite{331cs}.
 Thus, the neutral currents involving  $Z$ and $Z^\prime$ have the form
\begin{equation}
\label{nc}
\mathcal{L} = \frac{g}{2\cos\theta_W} \overline{q}\gamma^{\mu}\left[\left(g_{_V}^{q}+g_{_A}^{q}\gamma_{5}\right)Z_{\mu} + \left(f_{_V}^{q}+f_{_A}^{q}\gamma_{5}\right)Z_{\mu}^{\prime}\right]q
\end{equation}
The respective vector, $g_{V}^{q}$, $f_{V}^{q}$, and axial, $g_{A}^{q}$, $f_{A}^{q}$, couplings are shown in Appendix.

\section{Production of the $J_3$ Leptoquark and Discovery Analysis}

\begin{figure}
\includegraphics[scale=0.3]{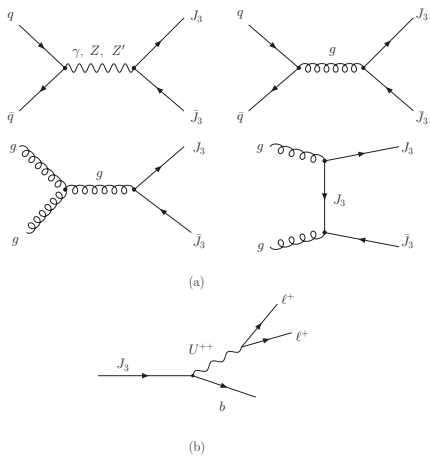}
\caption{The Feynman diagrams for  (a) the exotic quark pair production from Drell-Yan processes and QCD interactions, and (b) its decays involving the final particles.}
\label{fig:1}
\end{figure}

\begin{figure}
\resizebox{0.30\textwidth}{!}{%
\includegraphics[height=.20\textheight]{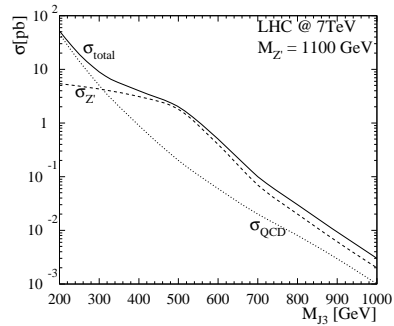}}
\caption{Total cross section and the partial contributions from the $Z^\prime$ and QCD for the $J_3$ pair
 production at the 7 TeV LHC.}
\label{fig:sc}
\end{figure}

We now consider the leptoquark pair production, $J_3\bar{J}_3$,
at the 7 and 14 TeV LHC. The $J_3\bar{J}_3$ production occurs through two main processes via gauge interactions: (1) the Drell-Yan process, involving the intermediation of the photon, and the neutral gauge bosons $Z$, $Z^{\prime}$, and (2) the QCD interactions, as shown in Fig.~\ref{fig:1}.
\noindent
In  Fig.~\ref{fig:sc} we show the total $J_3\bar{J}_3$  production cross section, $\sigma_{total}$, as a sum
of  Drell-Yan via $Z^{\prime}$, $\sigma_{Z^\prime}$, and QCD contributions $\sigma_{QCD}$, for a fixed $Z^\prime$ mass of $1.1$ TeV. Cross sections of order of picobarns are typical for $J_3$ masses below $\sim 500$ GeV. Even for TeV masses, cross sections in the femtobarn range are expected at the 7 and 14 TeV LHC.

For a light $J_3$ the major contribution comes from QCD interactions, due the large initial state gluon luminosity. However, as the $J_3$ mass increases, the $\sigma_{Z^\prime}$ contribution turns out to be dominant, as the gluon luminosity decreases steeply. This is also the outcome of a very peculiar feature of the model -- the $J_3$ coupling to the $Z^\prime$ is large due a factor $1/\sqrt{1-4\,\mathrm{sw^2}}$, as shown in the preceding section. Contributions from photons and $Z$ bosons are very small compared to the $Z^\prime$ contribution.

\begin{figure}
\resizebox{0.34\textwidth}{!}{%
\includegraphics[height=.22\textheight]{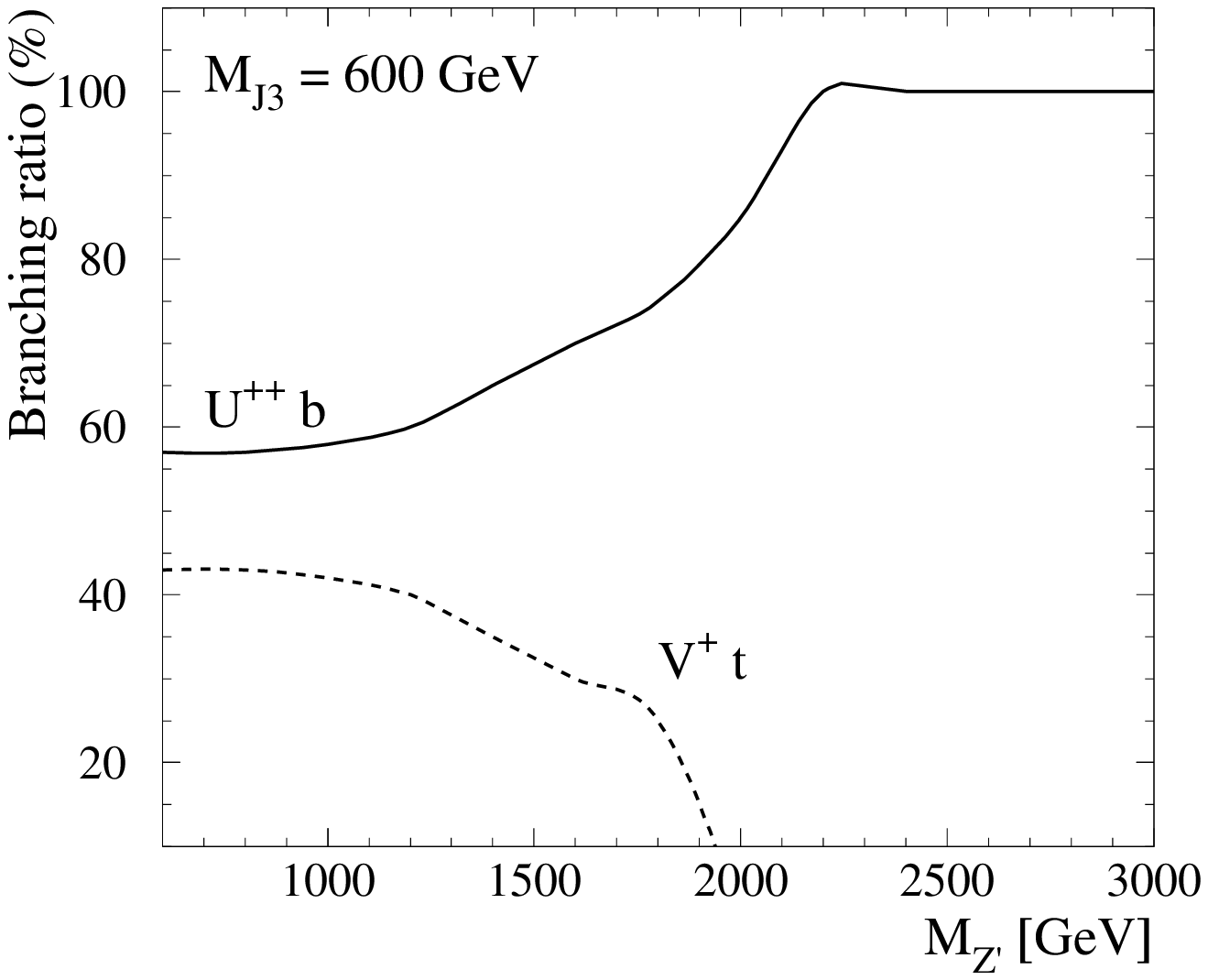}}
\resizebox{0.34\textwidth}{!}{%
\includegraphics[height=.22\textheight]{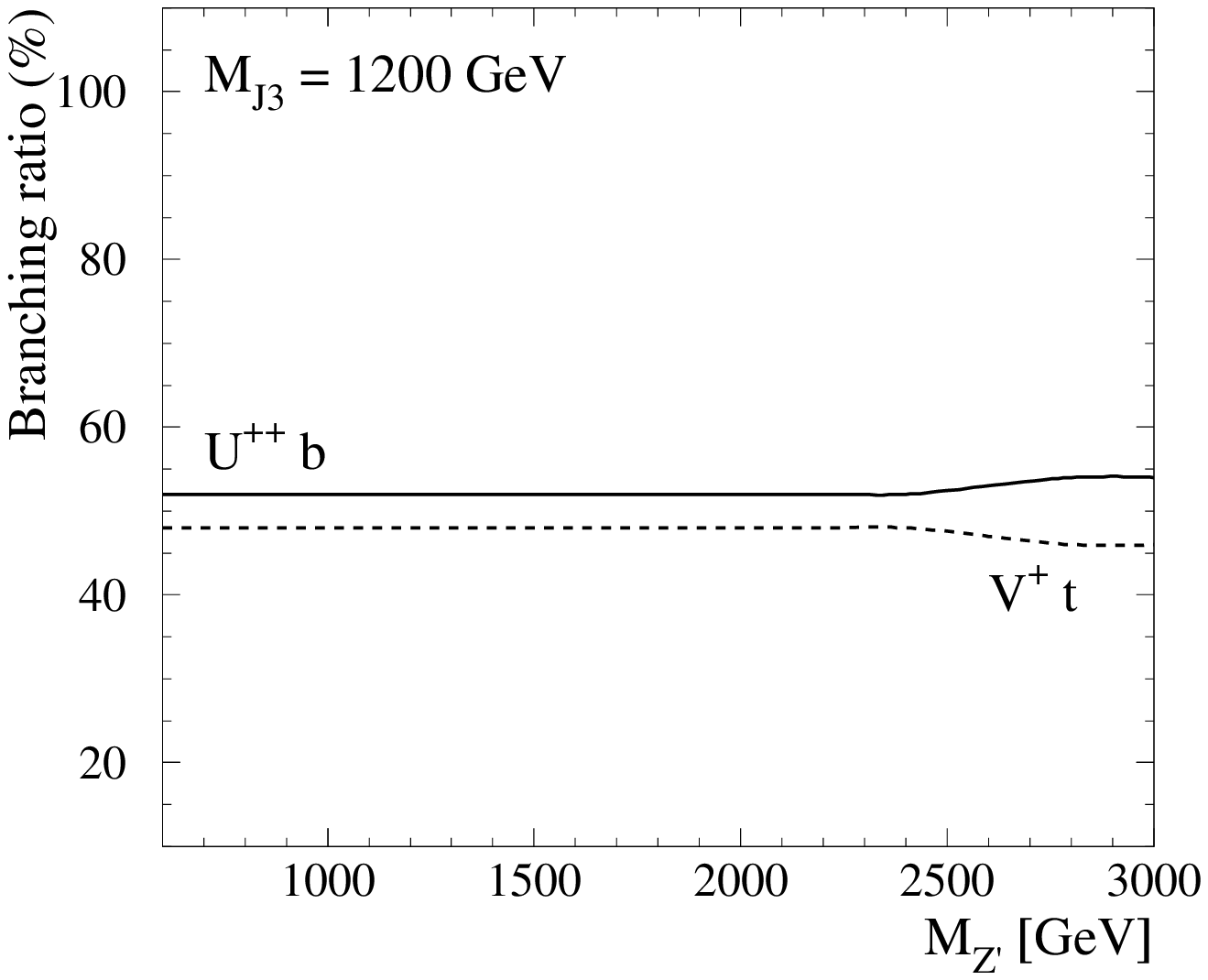}}
\caption{The branching ratios for the exotic quark  $J_3$ decay into $Vt$ and $Ub$ channels as function of the $Z^\prime$ mass. In the upper(lower) panel we show the branching ratios for a fixed $M_{J3}$ of $600$($1200$) GeV.}
\label{fig:br}
\end{figure}

Besides the decay involving the double charged bilepton plus bottom quark,  $J_3$  has another possible decay channel: the singly charged bilepton $V$ plus a top quark. We show in  Fig.~\ref{fig:br} the branching ratios of
 $J_3\rightarrow Vt$ and $J_3\rightarrow Ub$ decays as a function of ${Z^\prime}$  mass, fixing the $J_3$ mass. The channel $J_3\rightarrow Ub$ dominates from small to large leptoquark masses. The $Ub$ channel leads to a fully reconstructable final state presenting sharp resonances in the same-sign dileptons channel, associated to the $U$ decay, and in the same-sign dilepton plus a bottom jet, from the $J_3$ decay. Detecting these events is decisive to discriminate between competing models presenting exotic quarks.

Finally, we call the attention to the fact that the leptoquark  $J_3$ can couple also to single and doubly charged scalars allowing additional decay channels. However, these new branching ratios are very tiny (around 1\%) because the couplings between $J_3$ and the scalars are proportional to the  ratio of the quark bottom mass and a combination of the vacuum expectation values of order 1 TeV. Thus, the new channels involving scalars do not contribute to our analysis for the fermionic leptoquarks. The decay to $Vt$ would give rise to at least two hard leptons, jets and missing energy after the top quark decays. Despite its very distinctive signature, it is not possible to reconstruct the $V$ four momenta due the escaping neutrinos.

The leptoquark signature that emerges from the $J_3$ pair production and decay into a $U^{\pm\pm}b(\bar{b})$ channel is
\begin{equation}
pp \rightarrow J_{3}\bar{J}_{3}  \rightarrow b\bar{b}\;\ell^+\ell^+\ell^{\prime -}\ell^{\prime -}\;\; .
\label{j3pd}
\end{equation}
We do not demand the tagging of the bottom quark jet, so we propose the search for the inclusive $\ell^+\ell^+\ell^{\prime -}\ell^{\prime -}$ ($\ell,\ell^\prime=e,\mu$) plus two hard jets events.
We use the {\texttt CompHep} package~\cite{comp}, adopting the CTEQ6L parton
distribution functions ~\cite{CETQ}, with renormalization/factorization scales fixed at the exotic quark mass, to compute the signal and 3-3-1 backgrounds cross sections.
The irreducible SM background, $\ell^+\ell^+\ell^{\prime -}\ell^{\prime -} jj$ ($\ell,\ell^\prime=e,\mu$), was computed with the {\texttt MadGraph/MadEvent} package~\cite{Mad} also adopting the CTEQ6L parton
distribution functions and fixing the renormalization/factorization scale at the partonic center-of-mass energy, $\sqrt{\hat{s}}$.

For trigger and identification purposes we adopted the following cuts on the final state particles
\begin{eqnarray}
\label{id}
&& p_{T} (\ell) \geq 20\; \hbox{GeV},\,\,\, p_{T} (j) \geq 30\; \hbox{GeV},\,\,\, |\eta_{\ell,\,j}|  \leq 2.5\;, \nonumber\\
&& \Delta\,R_{\ell\ell} \geq 0.2\;,\;\;\;\;\; \Delta\,R_{jj} \geq 0.4\;, \;\;\;\;\; \Delta\,R_{j\ell} \geq 0.4
\end{eqnarray}

The SM backgrounds are quite small after imposing this set of acceptance cuts at both  the 7 and the 14 TeV LHC. The main 3-3-1 backgrounds come from the production of a $Z^\prime$ decaying to $U^{++}U^{--}$ plus a pair of jets from QCD radiation or a neutral gauge boson decay. To further suppress these and the SM backgrounds we impose the following cut on the invariant mass of the hardest jets pair:
\begin{equation}
\label{mjj}
M_{jj} \geq 120\; \hbox{GeV}.
\end{equation}

After applying this cut, the SM backgrounds are negligible while the signal to 3-3-1 background ratio is huge. We performed all analysis at the partonic level and do not include any detector effects. No K factors for QCD corrections were included either. This approach is reliable
in this case once the signal identification poses no problems, due the almost complete lack of background events.
Notwithstanding, a complete study including hadronization and detector effects is desirable to evaluate the impact of the extra QCD radiation in
the identification of the resonant structure arising from the leptoquark decay.

\begin{figure}
\resizebox{0.34\textwidth}{!}{%
\includegraphics[height=.25\textheight]{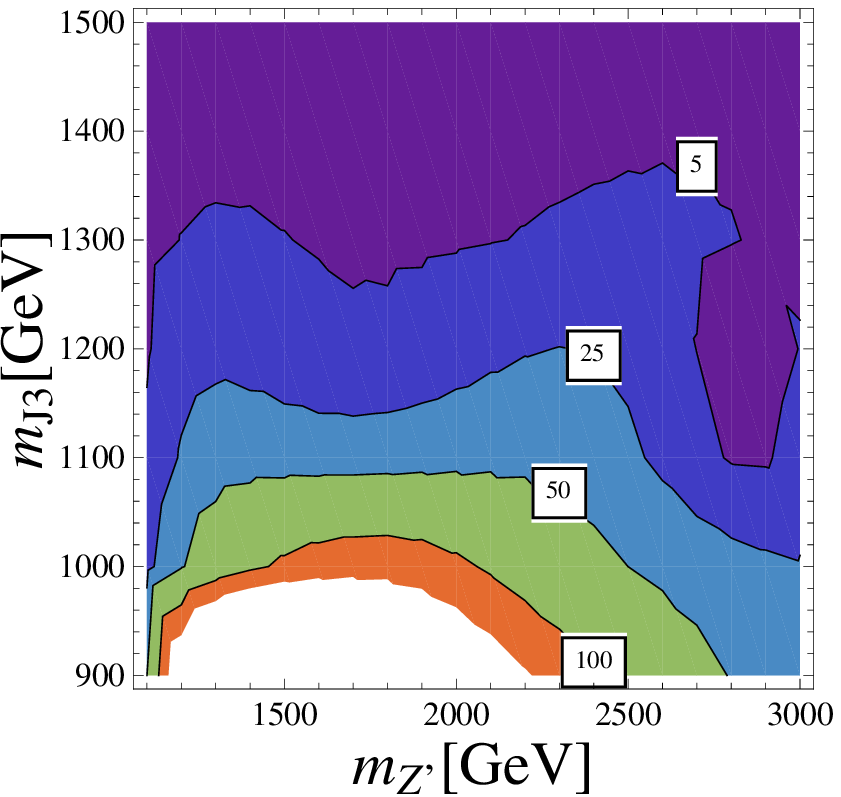}}
\resizebox{0.34\textwidth}{!}{%
\includegraphics[height=.25\textheight]{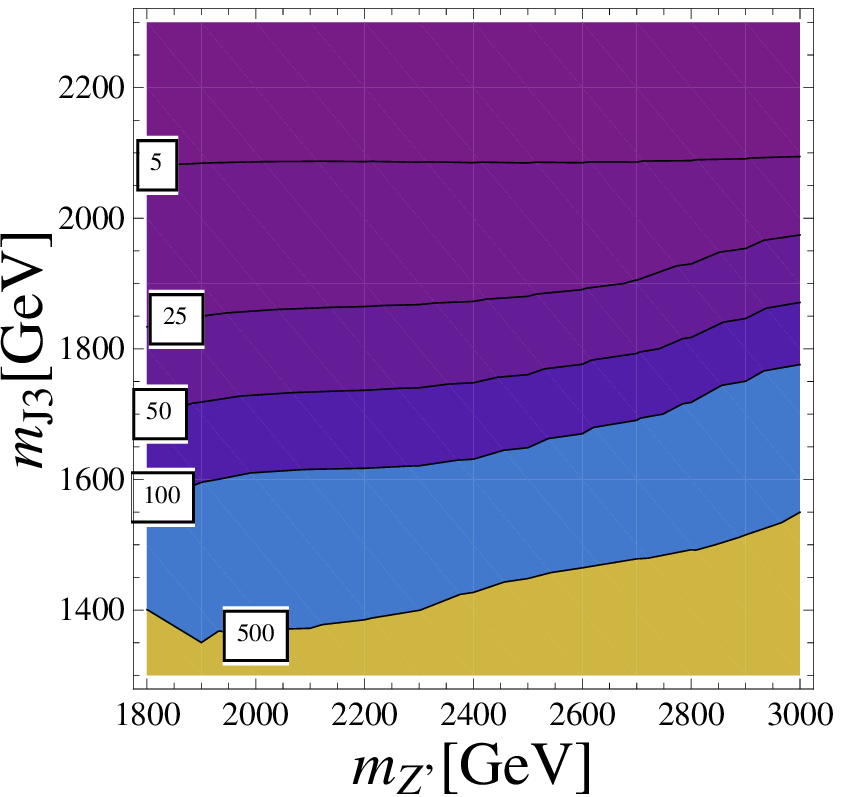}}
\caption{Upper panel: the contour lines represent the number of events expected for a fixed $5\; \hbox{fb}^{-1}$ of integrated luminosity at the 7 TeV LHC in the $m_{Z^\prime}$ {\it versus} $m_{J_3}$ plane. Lower panel: the same, this time for $20\; \hbox{fb}^{-1}$ and at the 14 TeV LHC. }
\label{mj3mzl}
\end{figure}

In the Fig. ~\ref{mj3mzl} we show the expected number of events for the signal  in the $m_{Z^\prime}$ {\it versus} $m_{J_3}$ plane for a fixed ${\cal L}=5\; \hbox{fb}^{-1}$ of integrated luminosity at the 7 TeV LHC (upper panel) and ${\cal L}=20\; \hbox{fb}^{-1}$ of integrated luminosity at the 14 TeV LHC (lower panel).

To understand the shape of the contour lines for the 7 TeV LHC, note that an increase in the number of events occurs along the direction of the slope $m_{Z^\prime}=2m_{J_3}$, where the leptoquarks are produced on their mass shell. On the other hand, as the cross section decreases very steeply with the $J_3$ mass for a fixed $Z^\prime$ mass, as we can see in Fig.~\ref{fig:sc}, the number of events drops quickly as the $Z^\prime$ becomes heavier. The left border of the plot can be understood, in its turn, as a consequence of the cut imposed on $M_{jj}$, which is bounded by $M_{jj}\leq\sqrt{m_{Z^\prime}^2-2m_U^2}\approx 0.9m_{Z^\prime}$. Because a candidate event must have $M_{jj}>120\gev$, the lighter the new neutral bosons are,  the smaller is the phase space left for hard jets production, which leads to a suppression in the number of events in this regime.

The lower panel shows the left border for the 14 TeV LHC actually. We see that as $m_{Z^\prime}$ increases, for a fixed $J_3$ mass, the number of events gets larger and will find a maximum for on shell production of $J_3$ pairs, the same behavior observed in the upper panel.

More interestingly, the Fig.~\ref{mj3mzl} shows that TeV masses are easily accessible at the LHC. The discovery of fermionic leptoquarks can be achieved up to $\sim 1.2\tev$ for a very broad mass range for $Z^\prime$ bosons. In the absence of events, $1.3$--$1.4\tev$ leptoquarks can be ruled out almost disregarded the $Z^\prime$ masses. Note that there would not be any other parameter dependence for the exclusion region. Placing limits in the  $m_{Z^\prime}$ {\it versus} $m_{J_3}$ plane would constrain the model in a straightforward manner.

At the 14 TeV LHC, the discovery region extends considerably, as can be seen at the lower panel of Fig.~\ref{mj3mzl}. With only $20\; \hbox{fb}^{-1}$, leptoquark masses up to $\sim 2\tev$ can be probed even for very heavy $Z^\prime$ bosons. Again, if no $b\bar{b}\;\ell^+\ell^+\ell^{\prime -}\ell^{\prime -}$ events are observed, multi-TeV limits can be imposed on the fermionic leptoquarks considered here.

It is shown in Ref.~\cite{h125331} that the 3-3-1 models can naturally explain the observed enhancement to diphoton decays of the  recently discovered boson at the 8 TeV LHC. In order to match the observed signal strength of diphoton signals within the $1\sigma$ error band, the new charged gauge bosons must have masses in the $\sim 200$--$800\gev$ range. In terms of the $Z^\prime$ masses, this corresponds to $\sim 715$--$2900\gev$ from Eq.~(\ref{mrel}) in  Appendix. As at the 8 TeV LHC we should expect a reach somewhat larger than that depicted in Fig.~\ref{mj3mzl}, we conclude that the doubly charged gauge boson, which would be also responsible for the diphoton enhancement, could be easily discovered for leptoquark masses up to $\sim 1.2\tev$ at the 8 TeV LHC.

We should point out that the Tevatron and LHC limits on $Z^\prime\rightarrow \ell^+\ell^-$~\cite{Zlinha} do not apply to our case
once the assumption of a narrow width does not hold. For example, for a 1100 GeV $Z^\prime$, the total width is $\Gamma_{Z^\prime} =$ 340 GeV.

\begin{figure}
\resizebox{0.34\textwidth}{!}{%
\includegraphics[height=.20\textheight]{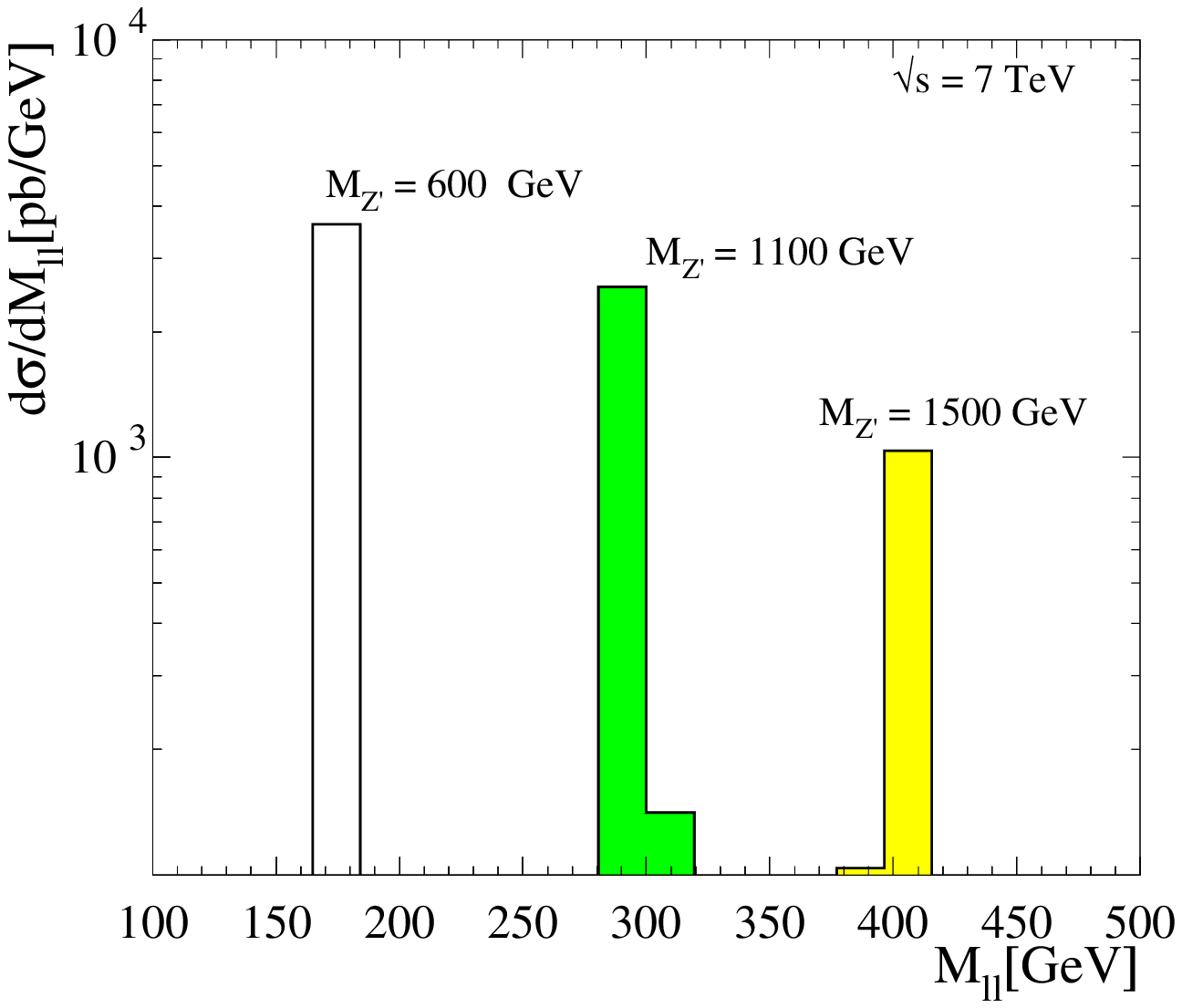}}
\resizebox{0.34\textwidth}{!}{%
\includegraphics[height=.20\textheight]{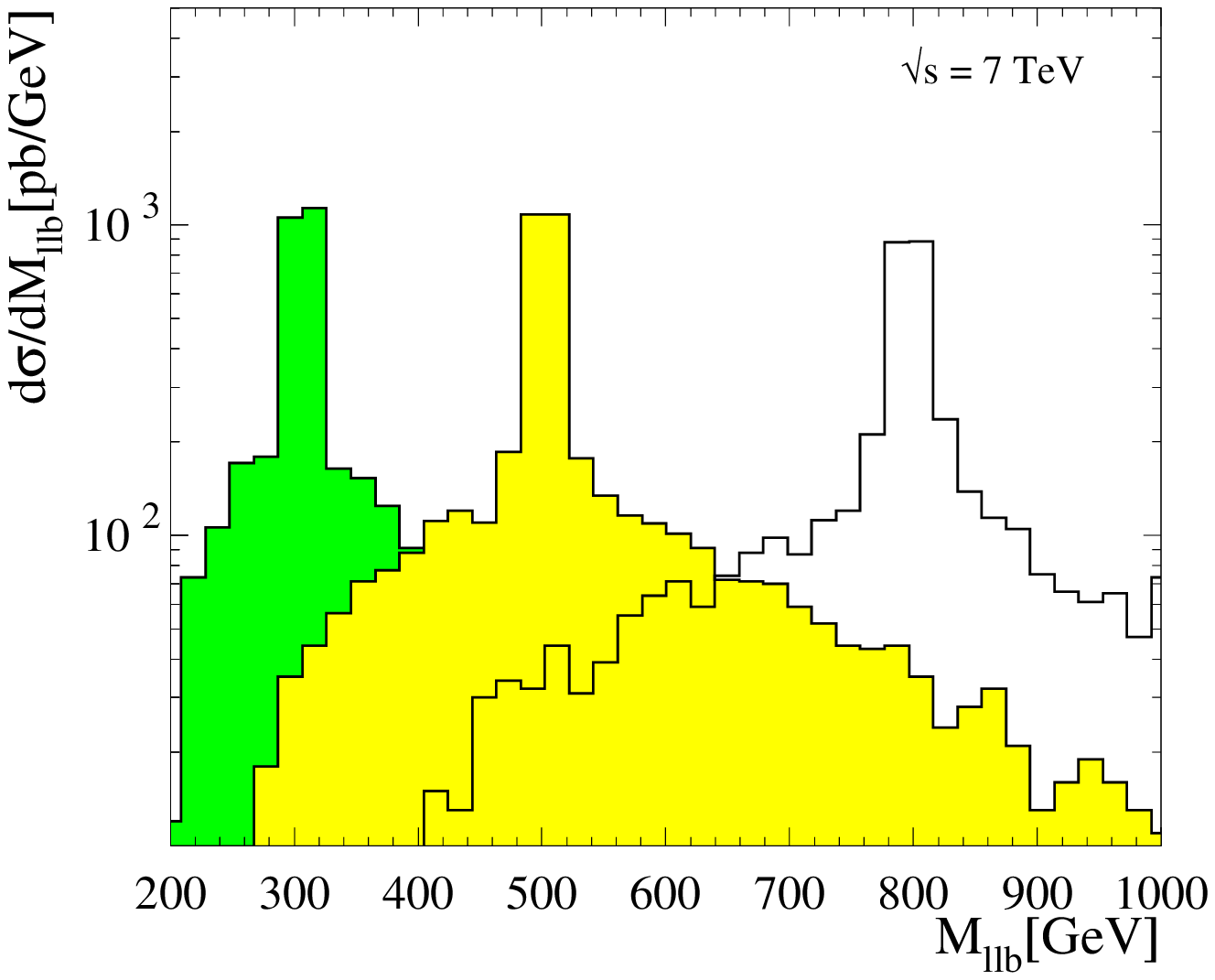}}
\caption{Upper panel: same-sign dilepton pair invariant mass distribution for three representative $Z^\prime$ masses at the 7 TeV
 LHC. Lower panel: invariant mass of a jet plus same-sign dileptons corresponding to a 300(500)(800) GeV $J_3$ mass. }
\label{invm}
\end{figure}

In the Figure~\ref{invm} we show the invariant mass distributions of a pair of same-sign dileptons, associated to the $U$ bilepton decay, and the jet plus same-sign dileptons, associated to the $J_3$ decay chain.
The sharp resonant structures in the $M_{\ell\ell}$ and $M_{j\ell\ell}$ invariant mass distributions can be clearly identified as a signal of the production and decay of fermionic  leptoquarks. Not only the exotic quarks, but also the doubly charged gauge boson shows itself in a tall sharp resonance.

Another important point to stress is the weaker model dependence of the $J_3$ decays compared to the models predicting scalar and vector leptoquarks.
The fermionic leptoquarks from the minimal 3-3-1 model have just two possible decay channels as discussed before. One of these involves the bilepton
gauge boson which, in its turn, decays exclusively into same-sign dileptons through gauge interactions. On the other hand, usual scalar and vector leptoquarks may decay to a pair of charged leptons and neutrinos, jets and charged and neutral leptons, and even to jet pairs, if they break baryon number conservation, as in the general R-parity-breaking supersymmetry~\cite{PDG}.

Moreover, in the case of scalar leptoquarks, a free Yukawa parameter is additionally involved in the leptoquark decay~\cite{CUY}, while the couplings of general vector leptoquarks to quark or lepton pairs may not be fixed by the gauge principle~\cite{CUY}.

It is beyond the scope of our work to make a detailed analysis comparing the discovery prospects for scalar, vector, and fermionic leptoquarks.
However, based on our results we believe it is much easier to establish the existence or not of fermionic leptoquarks compared to the spin-0 and spin-1 cases in a search for lepton-jet resonances.

\section{Conclusions}
 Fermionic leptoquarks and doubly charged gauge bosons are two types of exotic particles predicted in the minimal version of 3-3-1 models. The fermionic leptoquark $J_3$ is, in fact, an incarnation of a new heavy quark with electric
 charge $5/3$ carrying a nonvanishing lepton number.

The high production cross section of $J_3\bar{J}_3$ pairs, a high branching ratio of the leptoquark into same-sign dileptons plus a jet, and
an almost negligible level of background events associated to the final state topology, may lead to an immediate discovery of these new particles at the 7 TeV LHC even for masses of order of TeV. At the 14 TeV LHC the picture is even brighter, and a large portion of the parameter space of the model can be probed even for very heavy new states.

Based on our results, we believe it is much easier to establish the existence or not of fermionic leptoquarks as compared to their scalar and vector realizations of same mass, recalling that the production and decay of these fermionic leptoquarks occur, exclusively, through gauge interactions.

We finish by calling  attention to the large potential the LHC has to test  the channel for leptoquark decay we propose here. The experimental search for the signature we study in this work is worthwhile not only for probing the minimal 3-3-1 model, but also for constraining the way other gauge extensions of the SM for physics at the TeV scale can be constructed including or
not a vector bilepton and exotic quarks.

\acknowledgments
The authors thank FAPESP and CNPq for supporting this work.

\appendix

\section{Appendix\label{sec:ApA}}
The couplings for the ordinary quarks and the  $Z$ are  the same from the  SM. On the other hand,
the couplings between the new quarks and the $Z$  have only  vectorial character. Thus

\begin{eqnarray}
g_{V}^{j_1} & = g_{V}^{j_2}=\frac{3}{8} \mathrm{sw^2},\qquad g_{V}^{J_3}= -\frac{10}{3}\mathrm{sw^2}
\label{gj}
\end{eqnarray}
with $\mathrm{sw^2}=\mathrm{sin}^{2}\theta_{W}$
The  vector and the axial couplings between the u-type
quarks and the $Z^\prime$  are:
\begin{eqnarray}
& & f_{V}^{u}  =f_{V}^{c}=\frac{1-6\mathrm{sin}^{2}\theta_{W}}{2\sqrt{3-12\,\mathrm{sw^2}}},
\nonumber\\
& & f_{A}^{u}=f_{A}^{c}=\frac{1+2\mathrm{sin}^{2}\theta_{W}}{2\sqrt{3-12\,\mathrm{sw^2}}}
\nonumber\\
& & f_{V}^{t}=-\frac{1+4\,\mathrm{sw^2}}{2\sqrt{3-12\,\mathrm{sw^2}}}, \,\,\,\,
f_{A}^{t}=-\frac{1}{2\sqrt{3}}\sqrt{1-4\mathrm{sw^2}}
\label{fu}
\end{eqnarray}
and for the d-type quarks

\begin{eqnarray}
& & f_{V}^{d}  =f_{V}^{s}=\frac{1}{2\sqrt{3-12\,\mathrm{sw^2}}},\nonumber\\
& & f_{A}^{d}=f_{A}^{s}=\frac{1}{2\sqrt{3}}\sqrt{1-4\mathrm{sw^2}}\nonumber\\
& & f_{V}^{b}=-\frac{1-2\mathrm{sw^2}}{2\sqrt{3-12\,\mathrm{sw^2}}},
\,\,\,\,f_{A}^{b}=-\frac{1+2\mathrm{sw^2}}{2\sqrt{3-12\,\mathrm{sw^2}}}
\nonumber
\end{eqnarray}
\begin{eqnarray}
& &
f_{V}^{j_{m}}=-\frac{1-9\,\mathrm{sw^2}}{\sqrt{3-12\,\mathrm{sw^2}}},\,\,\,\,
f_{A}^{j_{m}}=-\frac{1-\mathrm{sw^2}}{\sqrt{3-12\,\mathrm{sw^2}}}
\nonumber\\
& & f_{V}^{J}=\frac{1-11\,\mathrm{sw^2}}{\sqrt{3-12\,\mathrm{sw^2}}},\,\,\,\,
f_{A}^{J}=\frac{1-\mathrm{sw^2}}{\sqrt{3-12\,\mathrm{sw^2}}}
\label{fd}
\end{eqnarray}

\section{Masses of the new gauge bosons \label{sec:ApB}}
For the new gauge bosons, $U^{\pm \pm}$, $V^{\pm}$,
and $Z^{\prime}$,  the following mass relation  is verified~\cite{331cs}:

\begin{equation}
M_{V} = M_{U} = \frac{\sqrt{3-12\,\mathrm{sw^2}}}{2\,\mathrm{cw}}\,M_{Z^{\prime}}
\label{mrel}
\end{equation}

\end{document}